\documentclass[preprint,superscriptaddress,nofootinbib]{revtex4-1}
\usepackage{graphicx, epsfig}

\usepackage{amsmath,amssymb}

\newcommand{\GeV}{{\text{GeV}}}

\begin{document}

\preprint{OU-HET-963}
\preprint{KANAZAWA-18-03}
\preprint{UT-HET 125}

\title{
Probing CP violating Higgs sectors via the precision measurement of coupling constants
}

%%%%%%%%%%%%%%%%%%%%%%%%
\author{Mayumi~Aoki}
\email{mayumi@hep.s.kanazawa-u.ac.jp}
\affiliation{
Institute for Theoretical Physics, 
Kanazawa University, 
Kanazawa 920-1192, Japan
}
%%%%%%%%%%%%%%%%%%%%%%%%
\author{Katsuya~Hashino}
\email{hashino@het.phys.sci.osaka-u.ac.jp}
\affiliation{
Department of Physics,
University of Toyama, 
3190 Gofuku, Toyama 930-8555, Japan
}
\affiliation{
Department of Physics,
Osaka University,
Toyonaka, Osaka 560-0043, Japan
}
%%%%%%%%%%%%%%%%%%%%%%%%
\author{Daiki~Kaneko}
\email{d$\_$kaneko@hep.s.kanazawa-u.ac.jp}
\affiliation{
Institute for Theoretical Physics, 
Kanazawa University, 
Kanazawa 920-1192, Japan
}
%%%%%%%%%%%%%%%%%%%%%%%%
\author{Shinya~Kanemura}
\email{kanemu@het.phys.sci.osaka-u.ac.jp}
\affiliation{
Department of Physics,
Osaka University,
Toyonaka, Osaka 560-0043, Japan
}
%%%%%%%%%%%%%%%%%%%%%%%%
\author{Mitsunori~Kubota}
\email{mkubota@het.phys.sci.osaka-u.ac.jp}
\affiliation{
Department of Physics,
Osaka University,
Toyonaka, Osaka 560-0043, Japan
}
%%%%%%%%%%%%%%%%%%%%%%%%

%%%%%% date %%%%%%%
%\date{\today}
%%%%%%%%%%%%%%%%%%%

\begin{abstract}

We study how effects of the CP violation can be observed indirectly by precision measurements of Higgs boson couplings at a future Higgs factory such as the international linear collider.
 We consider two Higgs doublet models with the softly broken discrete symmetry.
 We find that by measuring the Higgs boson couplings very precisely we are able to distinguish the two Higgs doublet model with CP violation from the CP conserving one.

\end{abstract}

\maketitle

%%%%%%%%%%%%%%%%%%%%%%%%%%
%%%  Introduction >>>  %%%
%%%%%%%%%%%%%%%%%%%%%%%%%%
\section{Introduction}

%%%%%%%%%%%%%%%%%%
%%%  Overview  %%%
%%%%%%%%%%%%%%%%%%
 By the discovery of a Higgs boson ($h$), the standard model~(SM) has been established as the low energy effective theory below the electroweak scale~\cite{Aad:2012tfa, Chatrchyan:2012xdj}. 
 In spite of such a success of the SM, we do not think that the SM is a fundamental theory because there are several phenomena which cannot be explained in the SM, such as baryon asymmetry of the universe~(BAU), dark matter, neutrino mass, cosmic inflation etc.
 Therefore an extension of the SM must be considered to describe these phenomena.
 This would be done at least partially by introducing an extended Higgs sector as seen in a promising scenario to explain BAU, the electroweak baryogenesis~\cite{Kuzmin:1985mm} where both additional CP violating phases and strongly first order electroweak phase transition (EWPT) can occur in an extended Higgs sector.

%%%%%%%%%%%%%%%%%%%%%%%%%%%%%%%%%
%%%% Measurements of CPV and Higgs boson coupling %%%%
%%%%%%%%%%%%%%%%%%%%%%%%%%%%%%%%%
Methods for exploring CP violating effects in extended Higgs sectors have been studied by the electric dipole moment~(EDM), angular distribution of $ h \to \tau^- \tau^+ $~\cite{DellAquila:1988bko, Berge:2008wi,  Jeans:2018anq} and the property of new particles via collisions between protons, photons or electron and positron~\cite{
Mendez:1991gp, Bernreuther:1993hq, Khater:2003wq, Asakawa:2000jy, Asakawa:2003dh, Keus:2015hva,Grzadkowski:2016lpv,Ogreid:2017alh,Bian:2017jpt,Chen:2017com,Belusca-Maito:2017iob,Basler:2017uxn,Azevedo:2018llq}.
 Meanwhile, we can test the strongly first order EWPT by measuring the SM-like Higgs coupling constants, especially the $hhh$ coupling which is  enhanced by several times 10~$\%$ from the SM prediction~\cite{
 Kanemura:2004ch,Noble:2007kk,Aoki:2008av,Aoki:2009vf,Aoki:2011zg,Kanemura:2011fy,Tamarit:2014dua,Hashino:2015nxa,Kanemura:2014cka,Kakizaki:2015wua,Hashino:2016rvx}. 
 The effects of the strongly first order EWPT can also be tested by detecting the characteristic spectrum of the gravitational waves which originate from the collision of the bubbles of the first order EWPT~\cite{
Kamionkowski:1993fg, Dolgov:2002ra, Grojean:2006bp,Espinosa:2008kw, Kehayias:2009tn, Kakizaki:2015wua,Caprini:2015zlo,Huber:2015znp,Hashino:2016rvx,Dev:2016feu,Chala:2016ykx,Huang:2016cjm,Kobakhidze:2016mch,Addazi:2016fbj,Hashino:2016xoj, Kang:2017mkl,Chao:2017ilw,Demidov:2017lzf,Chen:2017cyc,Chala:2018ari,Hashino:2018zsi,Vieu:2018zze,Bruggisser:2018mus,Wan:2018udw,Huang:2018aja,Bruggisser:2018mrt,Axen:2018zvb,Megias:2018sxv}.

%%%%%%%%%%%%%%%%%%%%%%
%%%%%%% In this latter %%%%%%%%
%%%%%%%%%%%%%%%%%%%%%%
In this letter, we examine how to indirectly detect the CP violating effects by precision measurements of the SM-like Higgs boson in two Higgs doublet models (2HDMs), where new CP violating effects can appear in the Yukawa couplings and in the Higgs potential.
 We focus on the 2HDM with a softly-broken $Z_2$ symmetry to avoid flavor changing neutral current~\cite{Glashow:1976nt}, which can contain a source of CP violation in the Higgs potential. 
 Under the symmetry the possible Yukawa couplings are classified in four types (Type-I, II, X and Y)~\cite{Barger:1989fj,Aoki:2009ha}.
 In the CP conserving case these types of Yukawa interaction can predict different patterns of deviations in the Higgs boson couplings, by which we are able to fingerprint each model if any of the deviation is detected in the couplings by precision measurements~\cite{Kanemura:2014dja, Kanemura:2014bqa, Kanemura:2018yai}.
 We here calculate the SM-like Higgs boson couplings to fermions and gauge bosons~($hff$ and $hVV$) in these 2HDMs with the CP violating phase. 
  The current data of the scaling factors for Higgs boson couplings by LHC are the following values: $\kappa_Z$ = 1.00 [0.92--1.00], $\kappa_W$ = 0.90 [0.81--0.99], $|\kappa_\tau| = 0.87^{+0.12}_{-0.11}$ and $|\kappa_b| = 0.49^{+0.27}_{-0.15}$ at $1\sigma$~\cite{Khachatryan:2016vau}.
 We here show how the effects of the CP violation can be indirectly observed by the precision measurements of the Higgs boson couplings at future collider experiments such as international linear collider~(ILC~\cite{Baer:2013cma,Asai:2017pwp,Fujii:2017vwa}, FCC-ee~\cite{Gomez-Ceballos:2013zzn}, CEPC~\cite{CEPC-SPPCStudyGroup:2015csa} and CLIC~\cite{CLIC:2016zwp}).

%%%%%%%%%%%%%%%%%%%%%%%%%%%%%%%%%%%%%%%%%%%%%%%%%%%%%%%%%%%%%%%%%%%%%%%%%
%%%%%%%%%%%%%%%%%%%%%%%%%%%%%%%%%%%%%%%%%%%%%%%%%%%%%%%%%%%%%%%%%%%%%%%%%

%%%%%%%%%%%%%%%%%%%%%%%%%%%%%%%%%%%%
%%%%%%%%%%% model (Potential) %%%%%%%%%%%%%%%%
%%%%%%%%%%%%%%%%%%%%%%%%%%%%%%%%%%%%
\section{2HDM with a softly-broken $Z_2$ symmetry }

We here introduce the 2HDMs with the softly broken discrete symmetry $Z_2$, which is introduced to avoid flavor changing neutral current~\cite{Glashow:1976nt}.
 Isospin doublet scalar fields $\Phi_1$ and $\Phi_2$ are transformed under the $Z_2$ symmetry: $\Phi_1\to\Phi_1$, $\Phi_2\to-\Phi_2$.
 The Higgs potential is given by
 %%%%%%
	\begin{align}
		V = &\mu_{1}^{2} (\Phi_{1}^{\dagger}\Phi_{1})
				+\mu_{2}^{2} (\Phi_{2}^{\dagger}\Phi_{2})
				-\{ \mu_{3}^{2} (\Phi_{1}^{\dagger}\Phi_{2})+h.c.\}
				+\frac{1}{2} \lambda_{1} (\Phi_{1}^{\dagger}\Phi_{1})^{2}
				+\frac{1}{2} \lambda_{2} (\Phi_{2}^{\dagger}\Phi_{2})^{2}
		\nonumber\\	&+\lambda_{3} (\Phi_{1}^{\dagger}\Phi_{1}) (\Phi_{2}^{\dagger}\Phi_{2})
					+\lambda_{4} (\Phi_{1}^{\dagger}\Phi_{2}) (\Phi_{2}^{\dagger}\Phi_{1}) 
					+\left \{ \frac{1}{2} \lambda_{5} (\Phi_{1}^{\dagger}\Phi_{2})^{2}+h.c. \right \} ,
	\label{eq:sofpot}
	\end{align}
%%%%%
where $\mu_3^2$ and $\lambda_5$ are generally complex, while the other parameters are real.
 $\Phi_1$ and $\Phi_2$ can be parameterised as 
%%%%%%
	\begin{align}
	\Phi_{1}=&
		\left(
		 	\begin{array}{c}
				w_{1}^{+} \\
				\frac{1}{\sqrt{2}} (v_{1}+h_{1}+i z_{1})
 			\end{array}
		\right)
	, \quad
	\Phi_{2}=
				\left(
		 	\begin{array}{c}
				w_{2}^{+} \\
				\frac{1}{\sqrt{2}} (v_{2}e^{i\xi}+h_{2}+i z_{2})
 			\end{array}
		\right),
	\end{align}
%%%%%%%
where $v^{2} \equiv (v_{1})^{2} +(v_{2})^{2} = (\sqrt{2} G_{F})^{-1} = (246~\GeV)^{2}$, $G_F$ being the Fermi coupling constant.
 In this paper, we use the redefinition of phases of doublet fields to absorb the $\xi$.
We then define the complex parameters $\mu_3^2$ and $\lambda_5$ as ${\rm Re}[\mu_3^2]+i{\rm Im}[\mu_3^2]$ and ${\rm Re}[\lambda_5]+i{\rm Im}[\lambda_5]$, respectively.

The stationary conditions are given by,
%%%%%%%%%%%%
	\begin{align}
	\frac{\partial V}{\partial h_{1}} \Bigg|_{\omega_j^+=h_j=z_j=0} =0 ,\quad
	 \frac{\partial V}{\partial h_{2}} \Bigg|_{\omega_j^+=h_j=z_j=0} =0 ,\quad
	 \frac{\partial V}{\partial z_{1}} \Bigg|_{\omega_j^+=h_j=z_j=0} =0\quad(j=1, 2),
	\end{align}
%%%%%%%%%%%%
which lead to the following equations:
%%%%%%%%%%%%
	\begin{align}
	\mu_{1}^{2}=\frac{M^{2}}{v^{2}} v_{2}^{2} -\frac{1}{2} ( \lambda_{1} v_{1}^{2}+\lambda_{345} v_{2}^{2}), \quad
	\mu_{2}^{2}=\frac{M^{2}}{v^{2}} v_{1}^{2} -\frac{1}{2} ( \lambda_{2} v_{2}^{2}+\lambda_{345} v_{1}^{2} ),\quad
		2 \mbox{Im}[\mu_{3}^{2}] = v_{1} v_{2} \mbox{Im}[\lambda_{5}],
		\label{eq:mv}
	\end{align}
%%%%%%%%%%%%
where $\lambda_{345}\equiv\lambda_3+\lambda_4+{\rm Re}[\lambda_5]$ and $M^2\equiv v^2\mbox{Re}[\mu_3^2]/v_1v_2$.
 There is one CP violating parameter in Higgs potential by using third equation in Eq.~(\ref{eq:mv}).
 In this letter, we treat ${\rm Im}[\lambda_5]$ as one physical parameter of CP violation.

We introduce the mixing angle $\beta$ ($\tan\beta=v_2/v_1$) in order to rotate the original basis to the Higgs basis~\cite{Davidson:2005cw}:
%%%%%%%%%%%%
 \begin{small}
	\begin{align}
	\left(
		 \begin{array}{c}
			\phi_{1}	\\
			\phi_{2}	
		\end{array}
	\right)=
	\left(
		\begin{array}{rr}
			\cos\beta & \sin\beta \\
			-\sin\beta & \cos\beta
		\end{array}
	\right)
	\left(
		 \begin{array}{c}
		 	\Phi_{1}	\\
			\Phi_{2}	
		\end{array}
	\right), \quad
	\phi_{1}=\left(
			 \begin{array}{c}
				G^{+} \\ 
				\frac{1}{\sqrt{2}}(v+h_{1}'+i G^{0})
			\end{array}
		\right)
			 ,\quad
	\phi_{2}=\left(
			 \begin{array}{c}
				H^{+} \\ 
				\frac{1}{\sqrt{2}}(h_{2}'+i h_3')
			\end{array}
		\right),
	\label{betarot}
	\end{align}
 \end{small}
%%%%%%%%%%%%
where $G^+$, $G^0$ are Nambu-Goldstone boson states. In this basis, the mass of $H^\pm$ is 
%%%%%%%%%%%%
	\begin{align}
	m_{H^{\pm}}^{2} &= M^{2} - \frac{1}{2} v^{2} (\lambda_{4}+\mbox{Re}[\lambda_{5}]).
	\end{align}
%%%%%%%%%%%%
The mass matrix for $h_1'$, $h_2'$ and $h_3'$ is not yet diagonalised, and takes the form: 
%%%%%%%%%%%%
	 \begin{footnotesize}
	\begin{align}
	\mathcal{M}^{2}=
	\left(
	\begin{array}{ccc}
		\tilde{m}_{h}^{2} \sin^2(\beta-\tilde{\alpha})+\tilde{m}_{H}^{2} \cos^2(\beta-\tilde{\alpha}) & \frac{1}{2}(\tilde{m}_{h}^{2}-\tilde{m}_{H}^{2}) \sin2(\beta-\tilde{\alpha}) & -\frac{1}{2}v^{2}\mbox{Im}[\lambda_{5}] \sin{2\beta} \\
		\frac{1}{2}(\tilde{m}_{h}^{2}-\tilde{m}_{H}^{2}) \sin2(\beta-\tilde{\alpha}) & \tilde{m}_{h}^{2} \cos^2(\beta-\tilde{\alpha})+\tilde{m}_{H}^{2} \sin^2(\beta-\tilde{\alpha}) & -\frac{1}{2}v^{2}\mbox{Im}[\lambda_{5}] \cos2\beta \\
		-\frac{1}{2}v^{2}\mbox{Im}[\lambda_{5}] \sin2\beta & -\frac{1}{2}v^{2}\mbox{Im}[\lambda_{5}] \cos2\beta & \tilde{m}_{A}^{2}
	\end{array}
	\right),
		\label{mt:mcpv}
	\end{align}
	 \end{footnotesize}
%%%%%%%%%%%%
where $\tilde{m}_h$, $\tilde{m}_H$ and $\tilde{m}_A$ are masses of the SM-like Higgs boson, extra CP-even and CP-odd Higgs bosons in the CP conserving limit, respectively. 
In this limit, $\tilde{\alpha}$ is the mixing angle which diagonalises two CP-even states in the Higgs basis. 
We use an orthogonal matrix $R$ in order to diagonalise the 3$\times3$ mass matrix in Eq.~(\ref{mt:mcpv}), 
%%%%%%%%%%%%
	\begin{align}
	R^{T} \mathcal{M}^{2} R={\rm diag}(m_{H_1}^{2},m_{H_2}^{2},m_{H_3}^{2}).
	\end{align}
%%%%%%%%%%%%
We treat the mass eigenstate $H_1$ as the (discovered) SM-like Higgs boson with the mass $125$ GeV. 
There are nine independent parameters in the potential in the following analysis:
%%%%%%%%%%%%
	\begin{align}
	v,M,m_{H^{\pm}},m_{H_1},\tilde{m}_{H},\tilde{m}_{A},\tilde{\alpha},\tan\beta,\mbox{Im}[\lambda_{5}].
	\label{para}
	\end{align}
%%%%%%%%%%%%

Next, we introduce Yukawa interactions and gauge interactions for $H_1$ in the model.
 Under the $Z_2$ symmetry, the Yukawa interaction is given by
\begin{align}
	-\mathcal{L}_{{\rm Yukawa}}= Y_{u}\bar{Q}_{L}(i \sigma_{2}\Phi_{u}^{*})u_{R}+Y_{d}\bar{Q}_{L}\Phi_{d}d_{R}+Y_{l}\bar{L}_{L}\Phi_{l}l_{R} +h.c.,
\end{align}
 where $\Phi_{u,d,l}$ are either $\Phi_{1}$ or $\Phi_{2}$ by the charge assignment of the $Z_2$ symmetry for fields in the model. There are 4 types of Yukawa interactions~\cite{Barger:1989fj,Aoki:2009ha} as shown Table~\ref{fg:z2assign}.
%%%%%%%%%%%%%%%%%%%%%%%
%%%%%%%%%%%%%%%%%%%%%%%
	\begin{table}[t]
	\begin{center}
		\caption{$Z_{2}$ charge assignment for fermions and scalar bosons in each Type~\cite{Aoki:2009ha}.}
	\begin{tabular}{|l||c|c|c|c|c|c|c|} \hline
				&$\Phi_{1}$	&$\Phi_{2}$	&$Q_{L}$	&$L_{L}$	& $u_{R}$	&$d_{R}$	& $l_{R}$	\\ \hline \hline
		Type-I	&$+$		&$-$			&$+$	&$+$	& $-$	& $-$	& $-$	\\ \hline
		Type-II	&$+$		&$-$			&$+$	&$+$	& $-$	& $+$	& $+$	\\ \hline
		Type-X	&$+$		&$-$			&$+$	&$+$	& $-$	& $-$	& $+$	\\ \hline
		Type-Y	&$+$		&$-$			&$+$	&$+$	& $-$	& $+$	& $-$	\\ \hline
	\end{tabular}
		\label{fg:z2assign}
	\end{center}
	\end{table}
%%%%%%%%%%%%%%%%%%%%%%%
%%%%%%%%%%%%%%%%%%%%%%%
Yukawa interactions for $H_1$ can be then rewritten as 
%%%%%%%%%%%%
	\begin{align}
	\begin{split}
		-\mathcal{L}_{{\rm Yukawa}}\ni& \sum_{f=u,d,l} \frac{m_{f}}{v}\bar{f} \left\{ (R_{11}+R_{21}\xi_{f})+(-2I_{f}) i \gamma_{5} R_{31} \xi_{f} \right\}f H_1,
	\label{higgsff}
	\end{split}
	\end{align}
%%%%%%%%%%%%
where $R_{ij}$ is the $(i, j)$ component of the matrix $R$, $\xi_f$, which depend on Type of 2HDM, are summarised in Table~\ref{tb:xi}, and $I_f$ is the third component of the isospin for fermion.

Gauge coupling constants to $H_1$ take the following form:
%%%%%%%%%%%%
	\begin{align}
		\mathcal{L}_{{\rm kin}}\ni R_{11}\left\{ \frac{2m_{W}^{2}}{v}W_{\mu}^+W^{-\mu}+\frac{m_{Z}^{2}}{v} Z_{\mu}Z^{\mu}\right\} H_1 
		= R_{11}\left\{ g_{hWW}^{\rm SM}W_{\mu}^+W^{-\mu}+\frac{1}{2}g_{hZZ}^{\rm SM} Z_{\mu}Z^{\mu}\right\} H_1.
	\label{higgsvv}
	\end{align}
%%%%%%%%%%%%
The scaling factors for $H_1VV$ ($V=W$ and $Z$) are given at the tree level by
	\begin{align}
		\kappa_{V} \equiv \frac{g_{H_1VV}}{g_{hVV}^{\rm SM}} = R_{11}.	
	\label{H1vv}
	\end{align}

%%%%%%%%%%%%%%%%%%%%%%%
%%%%%%%%%%%%%%%%%%%%%%%
	\begin{table}[b]
	\begin{center}
		\caption{$\xi_f$ factor for each Type~\cite{Aoki:2009ha}.}
	\begin{tabular}{|l||c|c|c|} \hline
		            & $\xi_u$ &$\xi_d$ & $\xi_l$ \\ \hline \hline
		Type-I  & $+\cot\beta$ & $+\cot\beta$ & $+\cot\beta$ \\ \hline
		Type-II & $+\cot\beta$ & $-\tan\beta$ & $-\tan\beta$ \\ \hline
		Type-X & $+\cot\beta$ & $+\cot\beta$ & $-\tan\beta$ \\ \hline
		Type-Y & $+\cot\beta$ & $-\tan\beta$ & $+\cot\beta$ \\ \hline
	\end{tabular}
		\label{tb:xi}
	\end{center}
	\end{table}
%%%%%%%%%%%%%%%%%%%%%%%
%%%%%%%%%%%%%%%%%%%%%%%

 There are the theoretical bounds on the parameter space in the 2HDM with the CP violation.
 The vacuum stability condition for the Higgs potential is given in Ref.~\cite{Grzadkowski:2009bt}.
 The perturbative unitarity bounds on the two-body elastic scattering amplitudes for the gauge and Higgs bosons are given in Refs.~\cite{Ginzburg:2005dt, Kanemura:2015ska}. 

 The constraints from the $S$, $T$ and $U$ parameters are seen in \cite{Peskin:1990zt, Peskin:1991sw, Haber:2010bw}.
 Parameters in the 2HDM are constrained by the direct searches of additional Higgs bosons by the data from LHC Run-1 and Run-2~\cite{Bernon:2015qea,Dorsch:2016tab,Han:2017pfo,Arbey:2017gmh,Chang:2015goa}.
 In addition, the flavour experiments such as $B$ meson decays give the lower limit on $m_{H^\pm}$ and $\tan\beta$ for each Type~\cite{Enomoto:2015wbn,Misiak:2017bgg}.
New CP violating effects in the new physics models are constrained by EDM. 
 The bounds from the EDM experiments on the parameter space of the 2HDM with CP violation have been discussed in Refs.~\cite{Abe:2013qla,Cheung:2014oaa}.

%%%%%%%%%%%%%%%%%%%%%%%%%%%%%%%%%%%%
%%%%%%%%%%% Numerical analysis %%%%%%%%%%%%%%%
%%%%%%%%%%%%%%%%%%%%%%%%%%%%%%%%%%%%

\section{Numerical analysis }

 In order to examine how CP violating phases in the Higgs sector affect the Higgs boson couplings, we evaluate the scaling factors $\kappa_V$ defined in Eq.~(\ref{H1vv}) and the ratio of the decay rate for $H_1\to f\bar{f}$ with identifying $H_1$ as the discovered Higgs boson with the mass of 125~GeV and the decay rate for $h\to f\bar{f}$ in the SM: 
 %%%%%%%%%%%%
	\begin{align}
	\frac{\Gamma_{\rm 2HDM}(H_1\rightarrow f\bar{f})}{\Gamma_{\rm SM}(h\rightarrow f\bar{f})}
	=&(c_f^s)^2+(c_f^p)^2\left(1-\frac{4m_f^2}{m_{H_1}^2}\right)^{-1},
		\label{eq:kappa}
	\end{align}
%%%%%%%%%%%%
where
%%%%%%%%%%%%
	\begin{align}
	c_{f}^{s}=R_{11}+R_{21} \xi_f,\quad c_{f}^{p}=(-2I_f) R_{31} \xi_f.
	\label{ctau}
	\end{align}
%%%%%%%%%%%%
The ratio of the decay rates coincides with that given in Ref.~\cite{Shu:2013uua}.
In the following numerical analysis, we take four of the nine parameters in Eq.~(\ref{para}) as 
%%%%%%%%%%%%
	\begin{align}
	v = 246~\GeV,\quad m_{H_1} = 125~\GeV,\quad \tilde{m}_{H} = 200~\GeV,\quad \tilde{m}_{A} = 250~\GeV.
			\label{VAL}
	\end{align}
%%%%%%%%%%%%
Since the ratio of the decay rates is independent of $M$ and $m_{H^\pm}$ at the tree-level, we can take values of $M$ and $m_{H^\pm}$ to avoid the current constraints from the $S$, $T$ and $U$ parameters~\cite{STUpdg}.
We show the numerical results for the scaling factor and the ratio of the decay rates varying the rest parameters~($\tan\beta$, $\tilde{\alpha}$ and ${\rm Im}[\lambda_{5}]$).
In the CP conserving limit, $\cos(\beta-\tilde{\alpha})$ correspond to $R_{21}$.

%----------- fig ---------------
	\begin{figure}
  \begin{center}
   \includegraphics[scale=0.58]{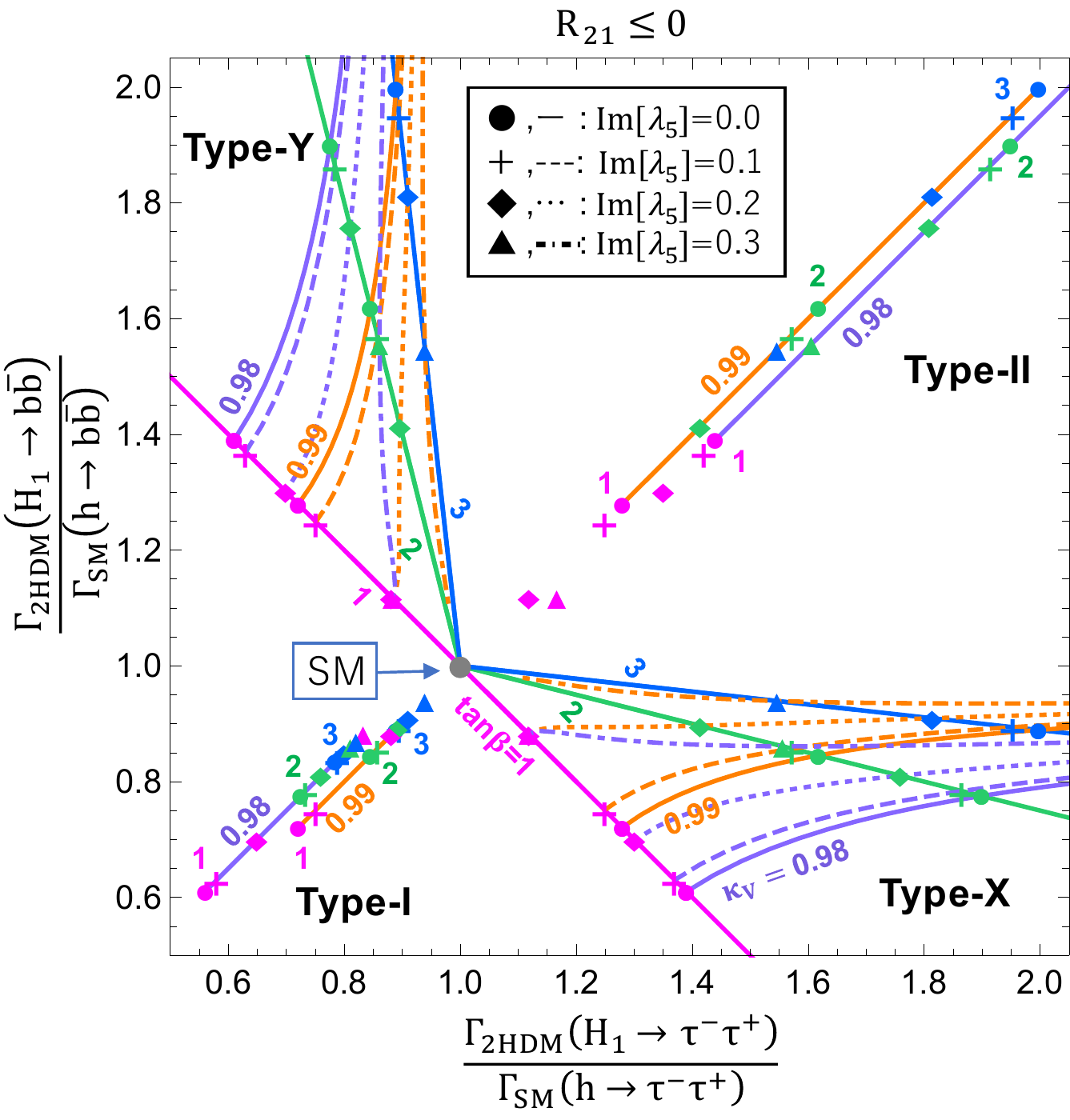}
   \includegraphics[scale=0.58]{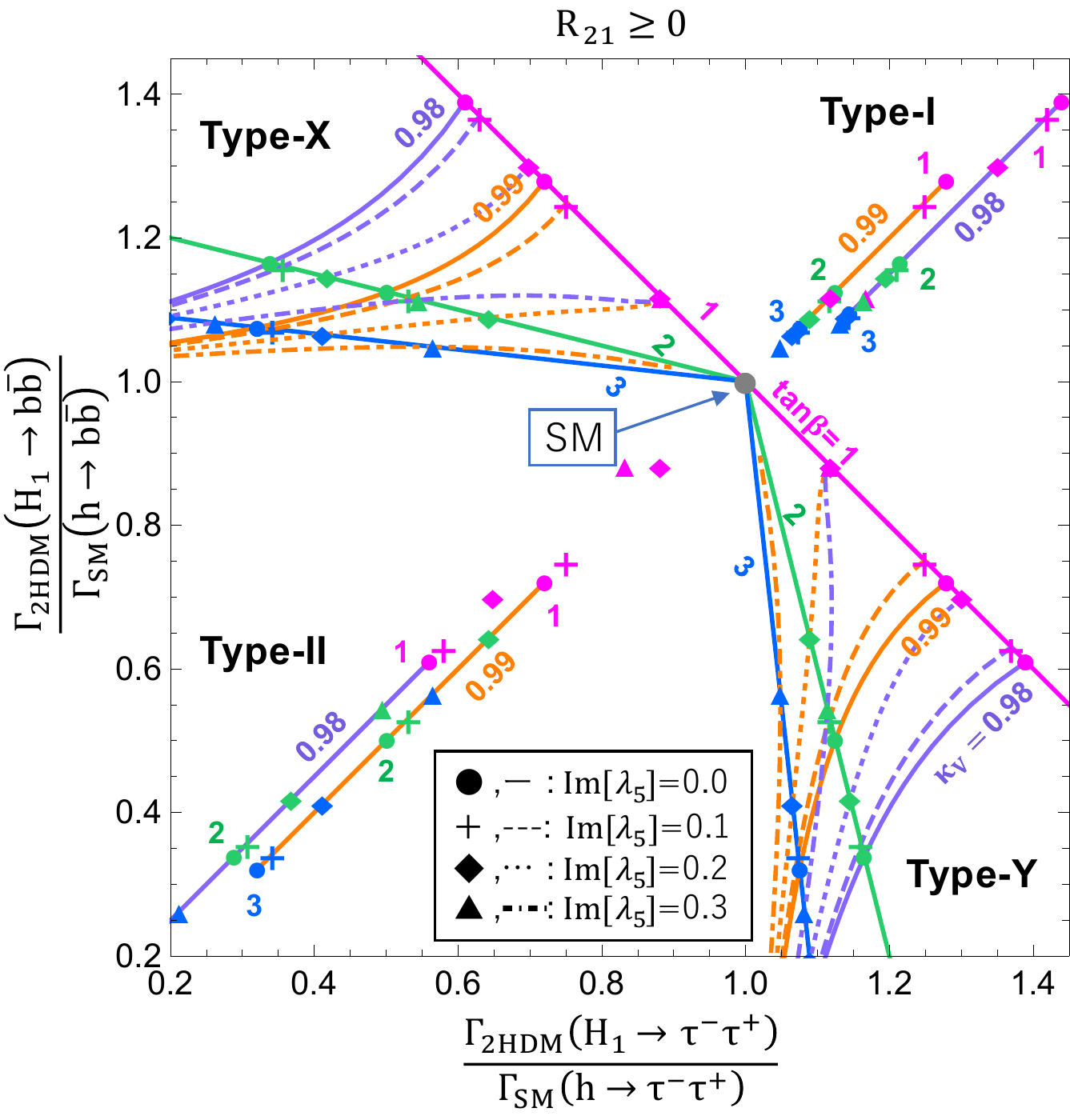}
   				\\[3mm]
   \includegraphics[scale=0.58]{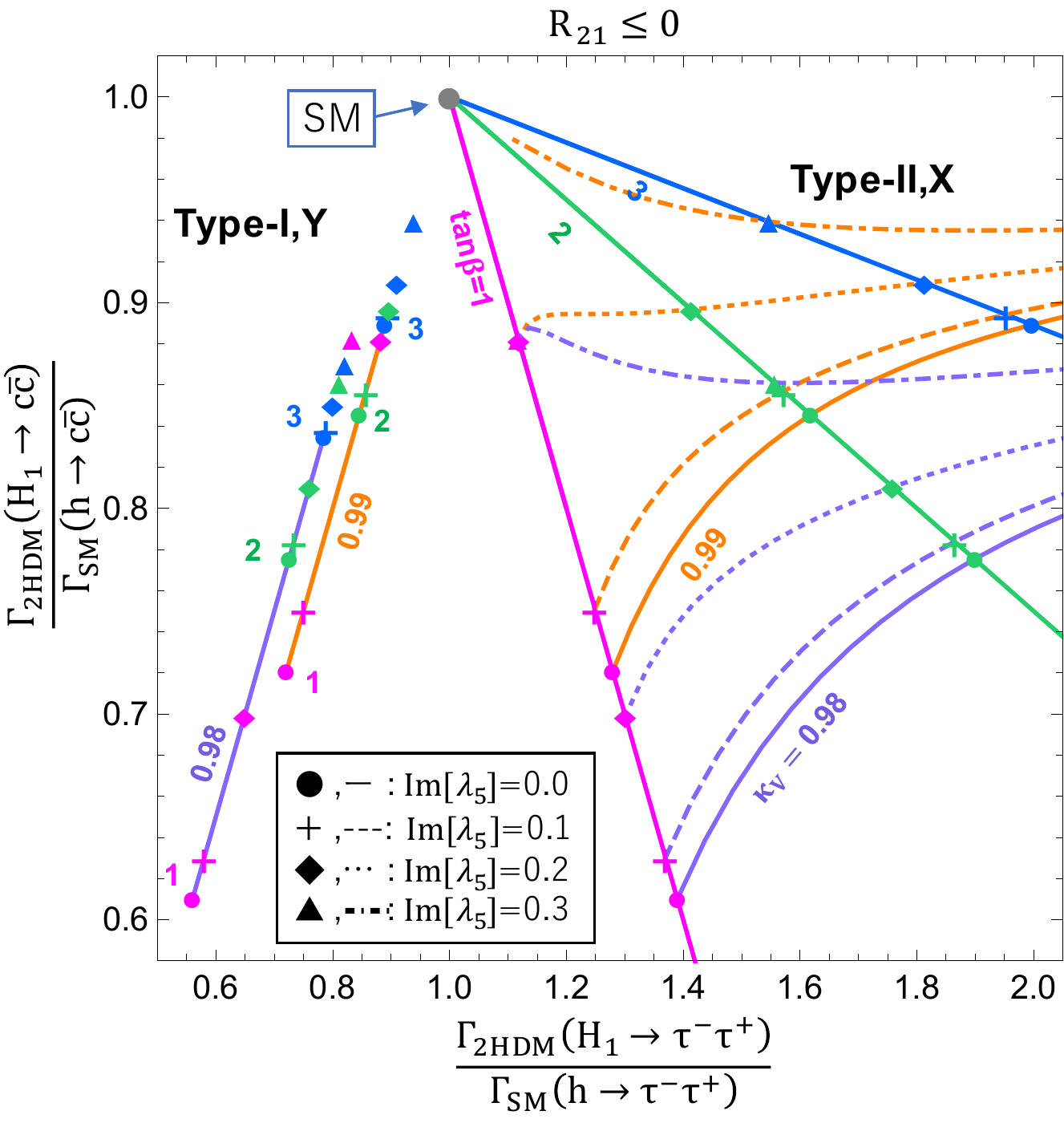}
   \includegraphics[scale=0.58]{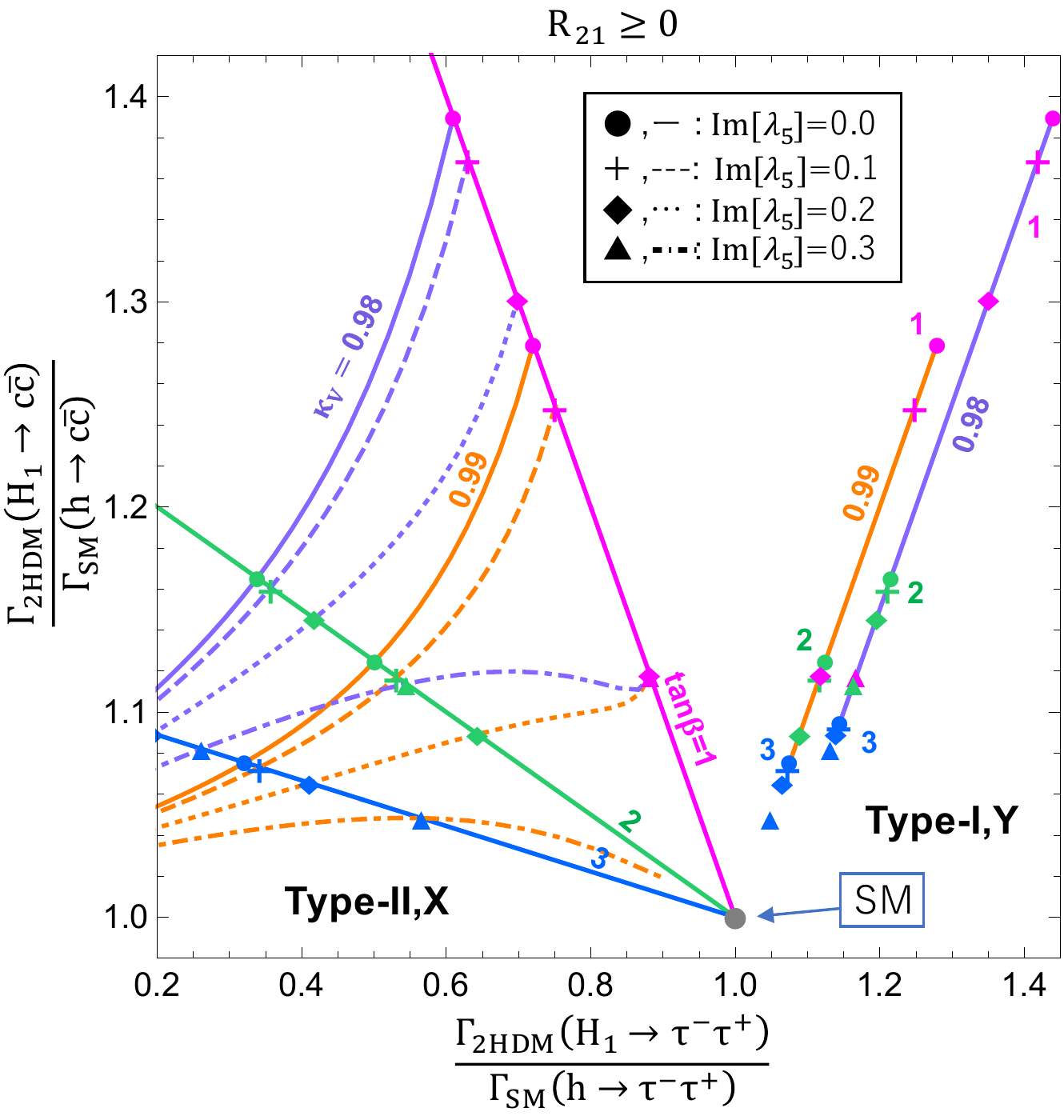}
   \vspace*{-5mm}
   \caption{ \small
 The ratio of decay rate for $H_1\to\tau^-\tau^+$, $H_1\to b\bar{b}$ (top) and $H_1\to c\bar{c}$ (bottom) with $R_{21}\leq0$ (left) and $R_{21}\geq0$ (right). 
 The purple and orange lines correspond to $\kappa_V$=0.98 and 0.99. The magenta, green and blue lines and points correspond to $\tan\beta$=1, 2 and 3.
 The dashed lines, dotted lines and dot-dashed lines correspond to ${\rm Im}[\lambda_5$]= 0.1, 0.2 and 0.3, respectively.
  The points of cross, rhombus and triangle respectively move from the points of circle for CP conserving case by ${\rm Im}[\lambda_5]=0.1, 0.2$ and $0.3$. The point of $(1.0, 1.0)$ corresponds to the SM.
  We shift purple lines of Type-I and II in the upper panel and those of Type-I and Y in the lower panel sideways, because the actual lines coincide with the orange lines.
}
   \label{SFfigure}
  \end{center}
	\end{figure}
%----------------------------------

In Fig.~\ref{SFfigure}, we show the ratio of decay rates for various final states of fermions in Type-I, II, X and Y 2HDM for several values of ${\rm Im}[\lambda_5$] ($=0.0, 0.1, 0.2$ and $0.3$).
In the upper panels, the results are shown on the plain of the decay into $\tau^-\tau^+$ and that into $b\bar{b}$, while in the lower panels, those on the plain of the decay into $\tau^-\tau^+$ and that into $c\bar{c}$ are shown.
In the left side panels, the results for $R_{21}\leq0$ are shown, while in the right side panels those for $R_{21}\geq0$ are shown.
In each figure, the point of $(1.0, 1.0)$ corresponds to the SM.
Values of $\tan\beta$ are taken to be $1, 2$ and $3$, and those of $\kappa_V$ are $0.99$ and $0.98$.
 The purple lines for Type-I and II in the upper panels and Type-I and Y in the lower panels are slightly moved sideways from the original positions, which coincide with orange lines.
 For each type of 2HDM, the purple (orange) solid, dashed, dotted and dot-dashed lines correspond to the cases with ${\rm Im}[\lambda_5]=0.0, 0.1, 0.2$ and $0.3$ for $\kappa_V = 0.98$ ($\kappa_V = 0.99$), respectively.
 For Type-X and Y in the upper panels and Type-II and X in the lower panels, the magenta, green and blue solid lines respectively correspond to $\tan\beta=1, 2$ and $3$.
 The points of cross, rhombus and triangle show how the predictions differ from the CP conserving cases marked as the circle points, where the cross, rhombus and triangle points correspond to ${\rm Im}[\lambda_5]=0.1, 0.2$ and $0.3$, respectively.
 The ratio of decay rates for various final states of fermions approaches $\kappa_V^2+(1-\kappa_V^2)\xi_f^2$ when ${\rm Im}[\lambda_5]$ increases.
  For $\tan\beta = 1$ $(\tan\beta=2)$ with $\kappa_V = 0.99$, the mass of $m_{H_1} = 125$ GeV cannot be realized for ${\rm Im} [\lambda_5] > 0.22$ (${\rm Im} [\lambda_5] > 0.28$).
 Therefore, the triangle points (${\rm Im} [\lambda_5] = 0.3$) for those cases are not shown in the figure, and the orange dot-dashed lines are broken at the points (with $\tan\beta\simeq2.2$) where $m_{H_1}$ cannot be 125 GeV.
 For the parameters of Eq.~(\ref{VAL}), we may be able to distinguish not only the Types of 2HDM~\cite{Kanemura:2014bqa} but also CP violating cases from CP conserving cases by the precision measurement of the Higgs boson couplings as seen in Fig.~\ref{SFfigure}.
 However, we cannot distinguish the ratios of decay rates with CP violating effects from those in the CP conserving 2HDM when $\tilde{m}_{A}$ is very large.

%----------- fig ---------------
	\begin{figure}[p]
  \begin{center}
    \includegraphics[scale=0.57]{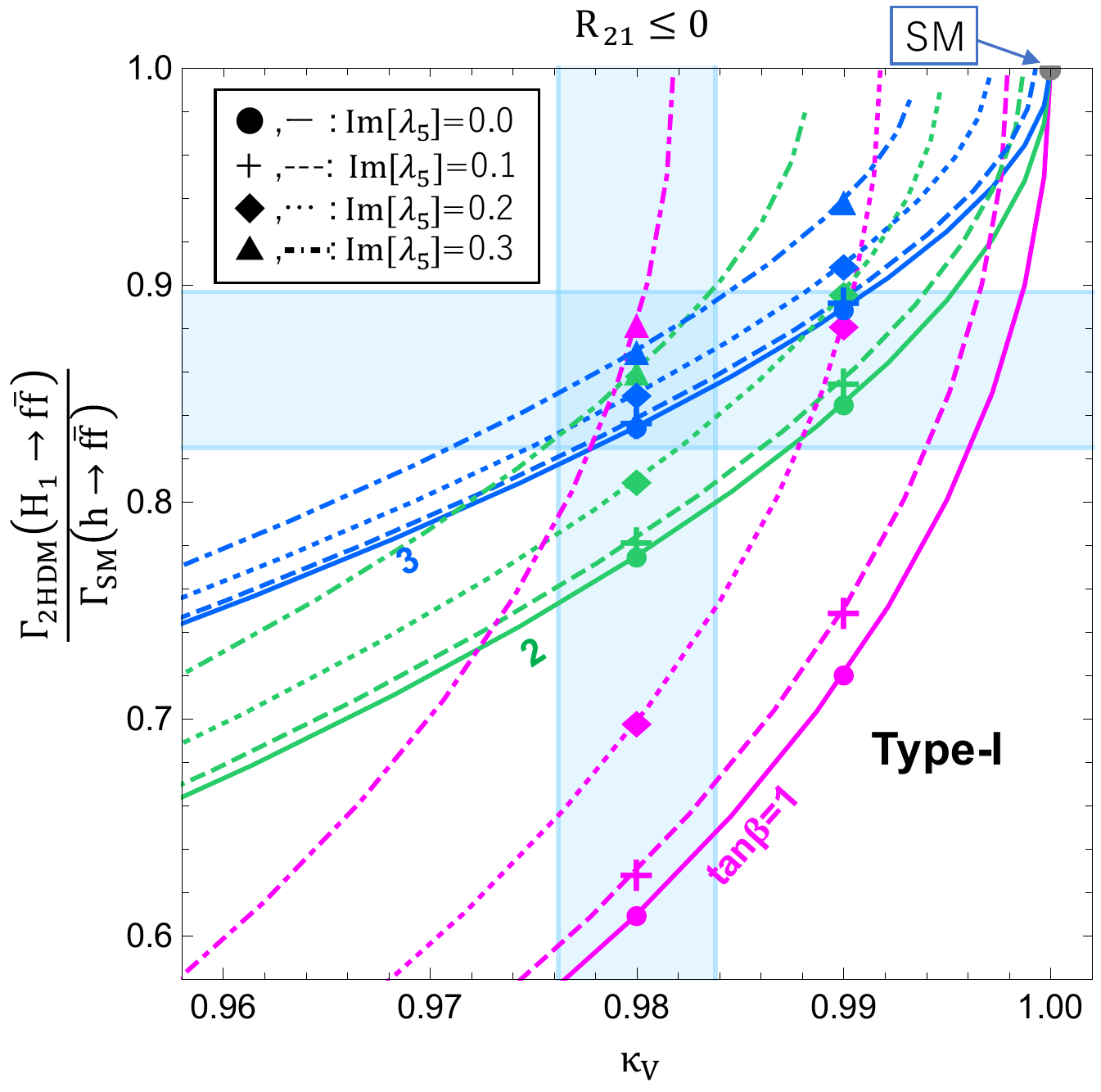}
    \includegraphics[scale=0.57]{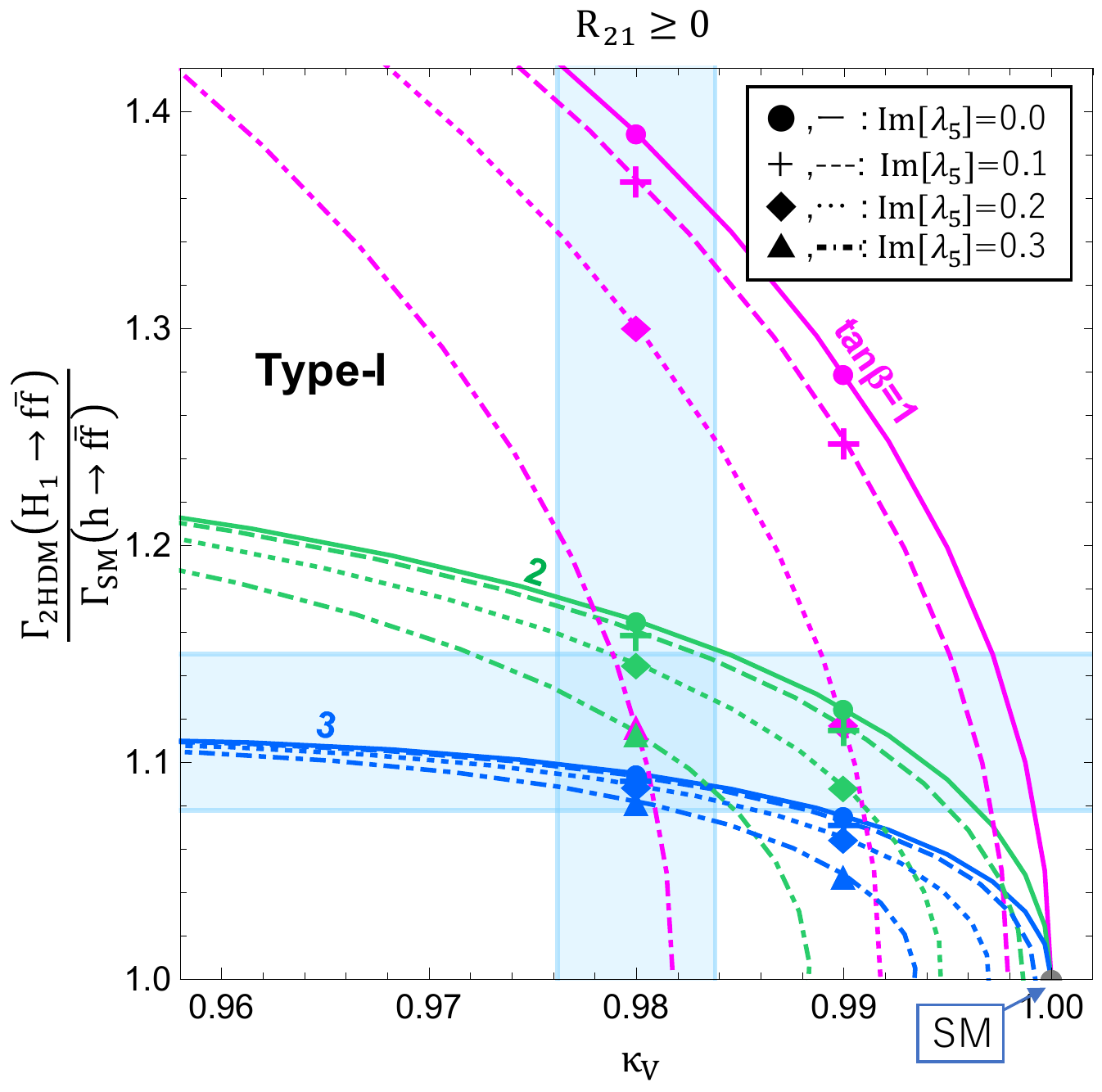}
    					\\ [3mm]
    \includegraphics[scale=0.58]{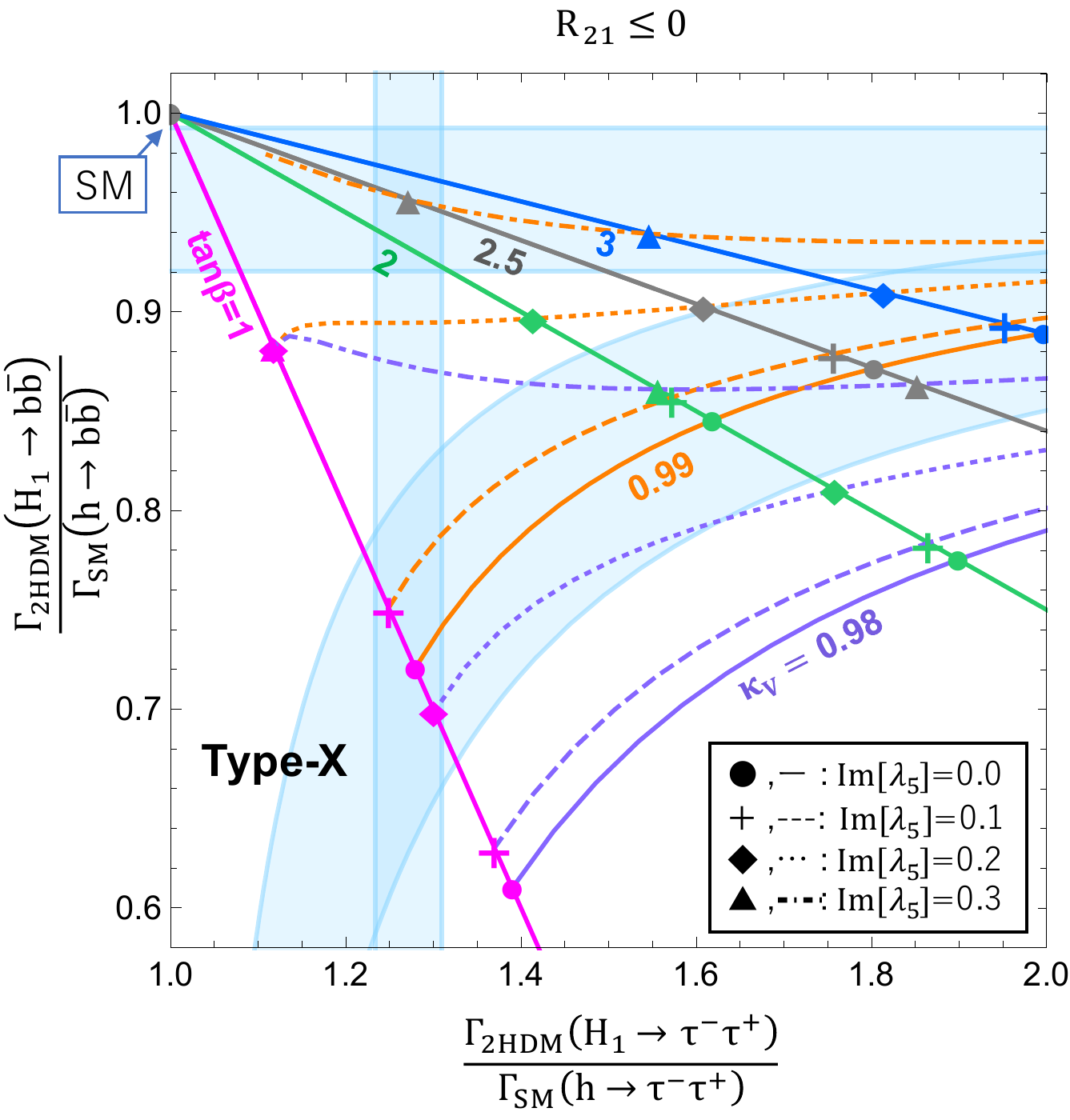}
    \includegraphics[scale=0.58]{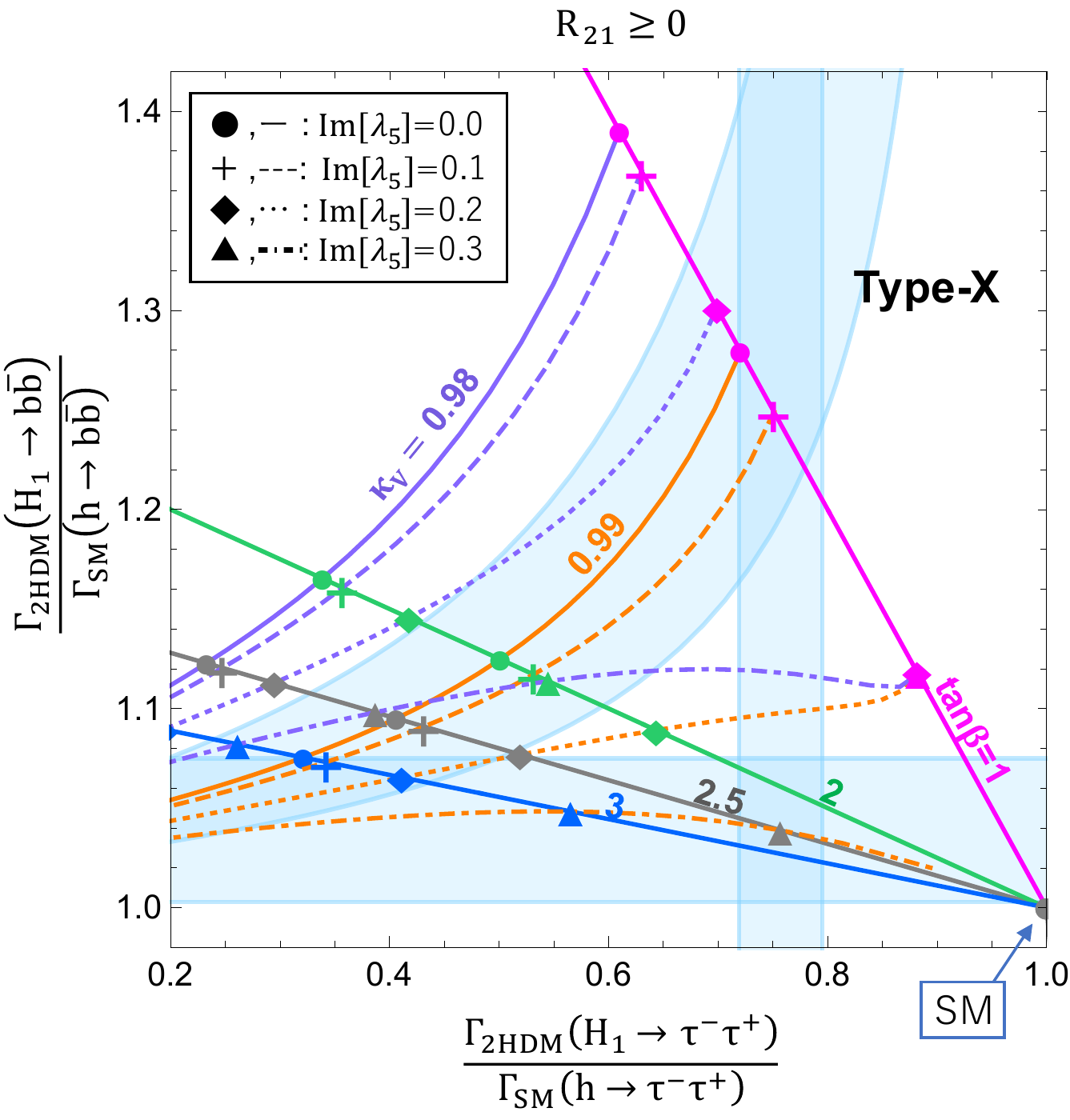}
    
   \caption{ \small
   The ratios of decay rate for the fermion and the scaling factors for the gauge boson in Type-I (top) and Type-X (bottom) with $R_{21}\leq0$~(left) and $R_{21}\geq0$~(right). 
   The grey-coloured lines in lower panel are $\tan\beta$=2.5.
   The other coloured lines, all kinds of lines and the marks are the same as those in Fig.~\ref{SFfigure}.
  The blue belts indicate the sensitivity region of the ILC with 250~GeV and ${\cal L}$=2 ab$^{-1}$ at the 1$\sigma$ accuracy \cite{Fujii:2017vwa}. The blue belt for $\kappa_V$ is the sensitivity region of $H_1ZZ$. In the upper panels, the blue belt for vertical axis is the sensitivity region of $H_1b\bar{b}$. 
  The centre values of the sensitivity for $H_1VV$ and $H_1f\bar{f}$ in the upper panels are at the green triangle point for $\kappa_V=0.98$.
  In the lower panels, the sensitivity for $H_1VV$ is along the orange solid line for $\kappa_V=0.99$, while the centre values of it for $H_1b\bar{b}$ and $H_1\tau^-\tau^+$ are at the grey triangle point for $\kappa_V=0.99$. 
   }
   \label{SFfigure2}
  \end{center}
	\end{figure}
%----------------------------------

\begin{figure}[t]
  \begin{center}
   \includegraphics[scale=0.57]{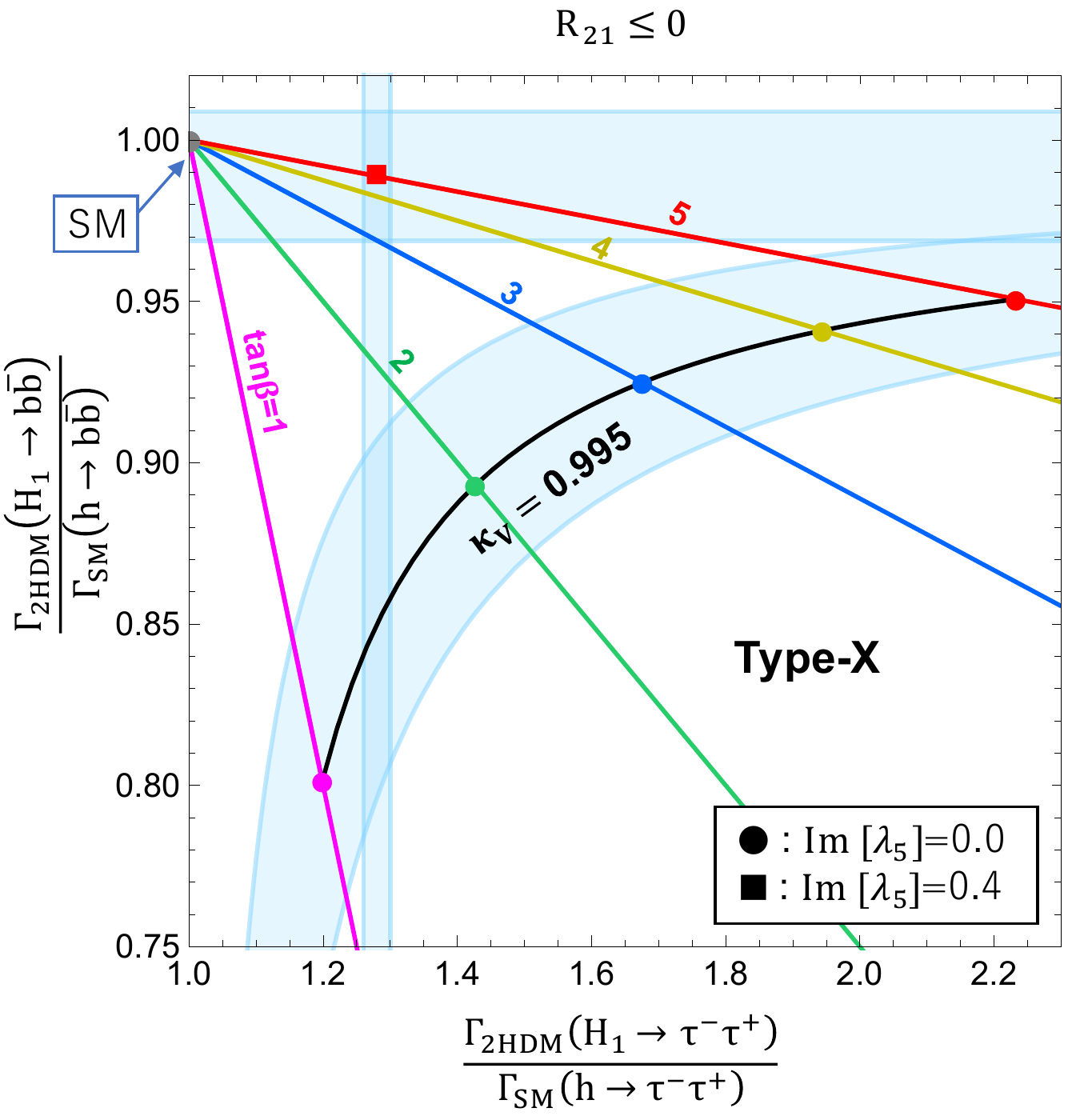}
   \includegraphics[scale=0.57]{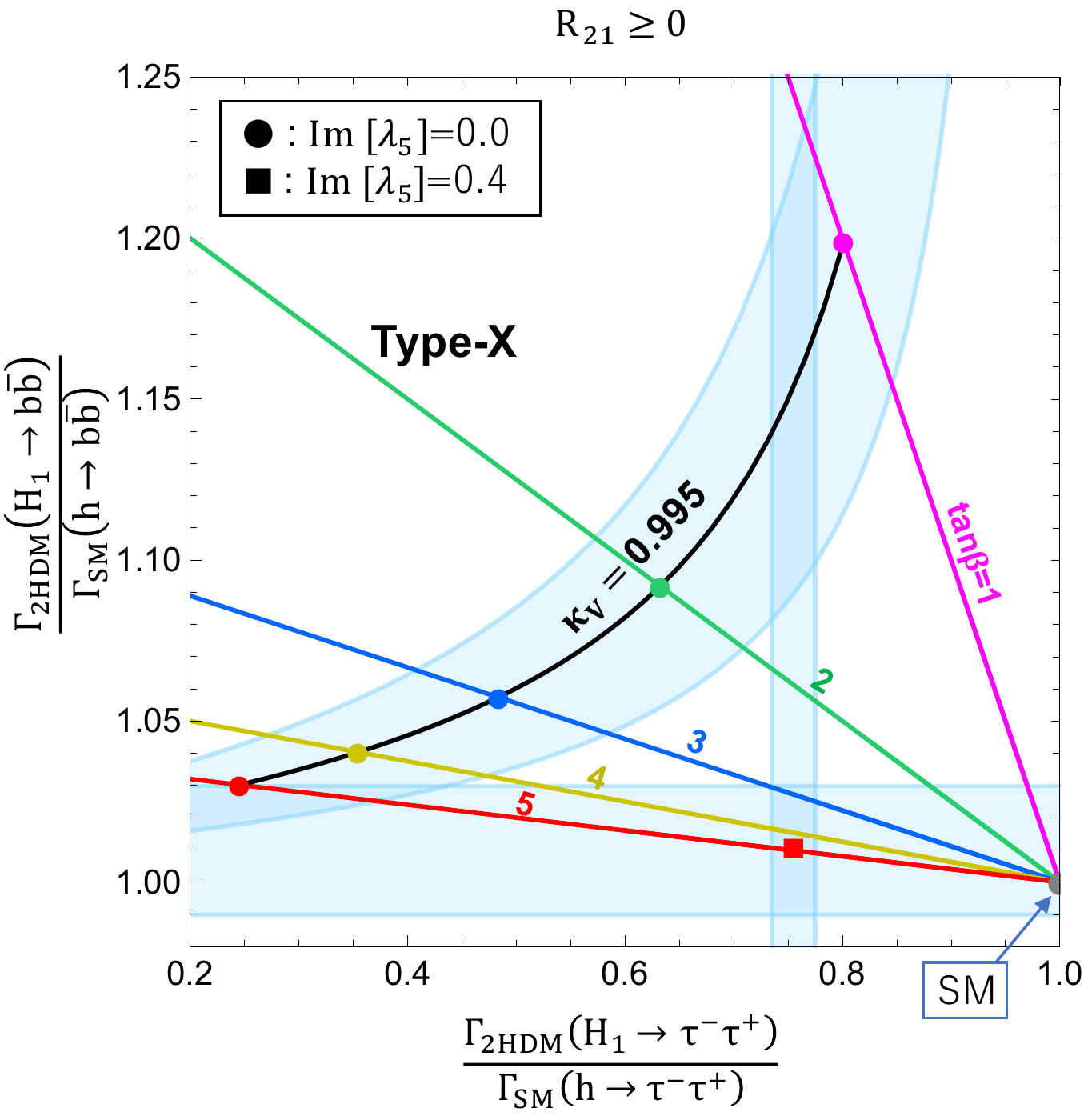}
   \caption{ \small
   The ratios of decay rate for the fermion and the scaling factors for the gauge boson in Type-X with $R_{21}\leq0$~(left) and $R_{21}\geq0$~(right).
   The magenta, green, blue, yellow and red solid lines and points in the figure respectively correspond to $\tan\beta=1, 2, 3, 4$ and $5$. All points correspond to $\kappa_V=0.995$. The red square point, which is not excluded by the EDM analysis based on Ref.~\cite{Cheung:2014oaa}, corresponds to ${\rm Im}[\lambda_5]=$ 0.4.
 The blue belts correspond to the sensitivity region for $H_1b\bar{b}$, $H_1\tau^-\tau^+$ and $H_1VV$, which respectively are about $1\%$, $1\%$ and $0.2\%$ precision at the 1$\sigma$ accuracy. The sensitivity for $H_1VV$ is along the black solid line for $\kappa_V=0.995$ and ${\rm Im}[\lambda_5]$=0, while it for $H_1b\bar{b}$ and $H_1\tau^-\tau^+$ is on the red square point for $\kappa_V=0.995$. 
  }
   \label{SFfigure5}
  \end{center}
	\end{figure}

In Fig.~\ref{SFfigure2}, we show whether we can distinguish the CP violating case from the CP conserving case by using the ILC with $\sqrt{s}=250$ GeV and ${\cal L}=2$ ab$^{-1}$. 
 We focus on the ratio of decay rates for $H_1\to f\bar{f}$ ($f=\tau, b$ and $c$) and the scaling factor $\kappa_V$ for $H_1VV$ in Type-I and X, 
because in Type-II and Y the parameters in Eq.(\ref{VAL}) are excluded by $b\to s\gamma$ \cite{Aoki:2009ha}.
In order to see how the CP violating case can be distinguished from the CP conserving case, we first do not take into account the EDM results in Fig.~\ref{SFfigure2}.
 Later in Fig.~\ref{SFfigure5}, the results where the EDM constraints are taken into account are shown.
 In the upper panels, the results in Type-I are shown on the plain of the ratio of decay rates for $H_1\to f\bar{f}$ ($f=\tau, b$ and $c$) and the scaling factor $\kappa_V$ for $H_1VV$, while in the lower panels, those in Type-X on the plain of the decay into $\tau^-\tau^+$ and that into $b\bar{b}$ are shown.
 In the left side panels, the results for $R_{21}\leq0$ are shown, while in the right side panels those for $R_{21}\geq0$ are shown.
 For Type-I 2HDM, the magenta (green and blue) solid, dashed, dotted and dot-dashed lines correspond to the cases with Im$[\lambda_5]=0.0, 0.1, 0.2$ and $0.3$ for $\tan\beta = 1$ ($\tan\beta = 2$ and $3$), respectively.
  For Type-X, the magenta, green, grey and blue solid lines respectively correspond to $\tan\beta=1, 2, 2.5$ and $3$.
 The points of cross, rhombus and triangle in Fig.~\ref{SFfigure2} are the same as those in Fig.~\ref{SFfigure}.
  As fiducial points, we take the green triangle point ($\tan\beta = 2$ and ${\rm Im}[\lambda_5] = 0.3$) with $\kappa_V=0.98$ in the upper panels, while we take the grey triangle point  ($\tan\beta = 2.5$ and ${\rm Im}[\lambda_5] = 0.3$) with $\kappa_V=0.99$ in the lower panels. Areas of the $1\sigma$ accuracy from the fiducial point are shown as the blue belts in the figures. 
Based on Ref.~\cite{Fujii:2017vwa}, we show the expected sensitivities for the future precision measurements of $H_1b\bar{b}$, $H_1c\bar{c}$, $H_1\tau^+\tau^-$ and $H_1ZZ$, which are taken to be $1.8\%$, $2.4\%$, $1.9\%$ and $0.38\%$ at the 1$\sigma$ accuracy, respectively.
  The belts for $H_1V V$ in the figure correspond to the sensitivity for $H_1ZZ$.
 The blue belts of the sensitivity for $H_1f\bar{f}$ in the upper panels correspond to the sensitivity for $H_1b\bar{b}$. 
In the lower panels the blue belts for $\kappa_V$ are taken for the orange solid lines ($\kappa_V = 0.99$ and ${\rm Im}[\lambda_5] = 0$). 
 
  In the upper panels, the fiducial points and the blue circle points with $\kappa_V=0.98$ are in the region where the blue belts of the sensitivity for $H_1f\bar{f}$ and $H_1V V$ overlap. 
In this case, we cannot distinguish the CP violating case from the CP conserving one in the Higgs sector by the precision measurements of Higgs boson couplings, unless $\tan\beta$ is determined accurately.  
In the CP conserving case for Type-X with $\kappa_V=0.99$, the ratios of decay rates for the fermion should be on the orange solid line.
However, the grey triangle for $\kappa_V=0.99$ is away from the blue belt of the sensitivity for $hVV$ in the lower panels. 
Therefore, we may be able to distinguish the CP violating case from the CP conserving case by the precision measurements of Higgs boson couplings.
 We note that the fiducial points in the figure are already excluded by the EDM.

%----------- fig ---------------
%----------------------------------

In the Type-I 2HDM being taken into account the EDM data, we confirmed that we cannot distinguish the CP violating case from the CP conserving case via the precision measurements of Higgs boson couplings,
because the ratios of decay rate for the fermion and the scaling factors for the gauge boson in these case overlap.
Therefore, in Fig.~\ref{SFfigure5}, we only show the results in the Type-X 2HDM under the constraint from the EDM data.
 In the left side panels, the results for $R_{21}\leq0$ are shown, while in the right side panels those for $R_{21}\geq0$ are shown.
The magenta, green, blue, yellow and red solid lines in the figure respectively correspond to $\tan\beta=1, 2, 3, 4$ and $5$.
 In the Type-X 2HDM $|c_u^p| < 3\times 10^{-2}$ with $c_u^p$ given in Eq.~(\ref{ctau}) is allowed by the EDM data~\cite{Cheung:2014oaa}.
 There is another constraint on ${\rm Im}[\lambda_5]$ with respect to satisfying the mass $125$ GeV of the Higgs boson.
In the Type-X 2HDM, if ${\rm Im}[\lambda_5]>0.41$ we cannot explain the Higgs mass $125$ GeV. 
Therefore in the figure the red square point (${\rm Im}[\lambda_5] = 0.4$) is taken as a fiducial point and the point is allowed by the EDM data.
The location of the red square point for the CP violating case is away from that of the red circle point for CP conserving case with the same values of the $\tan\beta$ and $\kappa_V$.
The blue belts in the figure correspond to the expected sensitivities for the future precision measurements of the Higgs boson couplings $H_1b\bar{b}$, $H_1\tau^-\tau^+$ and $H_1VV$, which are taken to be $1\%$, $1\%$ and $0.2\%$ at the $1\sigma$ accuracy.
 Such an accuracy could be achieved at the ILC with $\sqrt{s}=250$ GeV if the integrated luminosity is enhanced to be ${\cal L}=8$ ab$^{-1}$.
 In the figure, the blue belts for $H_1b\bar{b}$ and $H_1\tau^-\tau^+$ are on the red square point (${\rm Im}[\lambda_5]=0.4$), and the belt for the scaling factor $\kappa_V$ is along the black line for CP conserving case with $\kappa_V=0.995$.
  
Consequently, in the Type-I 2HDM it is difficult to distinguish the CP violating case from the CP conserving case by the very precise measurements of the Higgs boson couplings.   
  On the other hand, in the Type-X 2HDM we may be able to detect the CP violating effect by the very precise measurement of the Higgs boson couplings even in the case favoured by the EDM data, if the integrated luminosity is large enough.
  We note that in the Type-X 2HDM with $R_{21}\geq0$ we cannot distinguish the red square points for the CP violating case with ${\rm Im}[\lambda_5]=0.4$ from the points with $\tan\beta=$ 18--19, $\kappa_V=0.995$ and ${\rm Im}[\lambda_5]=0$.
However, in the CP conserving 2HDM, the case with such large $\tan\beta$ values with $\kappa_V=0.995$ is already excluded by current data \cite{CMS:2017pij}.

We here give a comment that the angular distribution of $H_1\to\tau^-\tau^+$ can be used to measure the CP violating effect in the Higgs sector~\cite{Jeans:2018anq}. 
The CP mixing angle $\psi_{CP}$ is given by
%%%%%%%%%%%%
	\begin{align}
	\mathcal{L}_{H_1\tau\tau}=g\bar{\tau}(\cos\psi_{CP}+i\gamma_{5}\sin\psi_{CP})\tau H_1,
	\end{align}
%%%%%%%%%%%%
where $g=-m_\tau\sqrt{(c_\tau^s)^2+(c_\tau^p)^2}/v$ with $c_\tau^s$ and $c_\tau^p$ given in Eq.~(\ref{ctau}).
At the ILC with $\sqrt{s}=250$ GeV and ${\cal L}=2$ ab$^{-1}$, $\psi_{CP}$ can be measured to a precision of $4.3^\circ$~\cite{Jeans:2018anq}.
In the Type-I 2HDM where are taken into account the EDM data, we cannot detect the CP violating effect by measuring the angular distribution of $H_1\to\tau^-\tau^+$ at the ILC.
On the other hand, in the Type-X 2HDM the corresponding values of $\psi_{CP}$ to the red square points in Fig.~\ref{SFfigure5} are given in Table.~\ref{tb:psiCP}.
We can complementarily examine the effects of the CP violation in the Type-X 2HDM by the precision measurements of the Higgs boson couplings and the angular distribution of $H_1\to\tau^-\tau^+$ at future Higgs factories.

%%%%%%%%%%%%%%%%%%%%%%%
	\begin{table}[t]
	\begin{center}
		\caption{The CP mixing angles $\psi_{CP}$ for the red square point in Fig.~\ref{SFfigure5}.}
	\begin{tabular}{|ll||c|c|} \hline
				&			& ${\rm Im}[\lambda_5]$ & $\psi_{CP}$ \\ \hline \hline
		Type-X	& ($R_{21}\leq 0$)	& $0.4$ & $-26^\circ$ \\ \hline
		Type-X	& ($R_{21}\geq 0$)	& $0.4$ & $-30^\circ$ \\ \hline
\end{tabular}
		\label{tb:psiCP}
	\end{center}
	\end{table}
%%%%%%%%%%%%%%%%%%%%%%%

%%%%%%%%%%%%%%%%%%%%%%%%%%%%%%%%%%%%
%%%%%%%%%%%      	  Summary		 %%%%%%%%%%%%
%%%%%%%%%%%%%%%%%%%%%%%%%%%%%%%%%%%%

\section{Summary}

We have studied how effects of the CP violation can be observed indirectly by precision measurements of the coupling constants of the Higgs boson with the mass 125 GeV at a future Higgs factory such as the ILC.
 We have investigated the difference between CP conserving and CP violating cases of the 2HDMs with the softly broken discrete symmetry.
 We have found that in some parameter sets the CP violating effects in the extended Higgs sectors can be detected by measuring the Higgs boson couplings very precisely.

%%%%%%%%%%%%%%%%%
%%%  Acknowledgments  %%%%
%%%%%%%%%%%%%%%%%

\begin{acknowledgments}
  The work of M.~A. is supported in part by the Japan Society for the
Promotion of Sciences (JSPS) Grant-in-Aid for Scientific Research (Grant
No. 25400250 and No. 16H00864). 
K.~H. and M.~K. are supposed by the Sasakawa Scientific Research Grant from The Japan Science Society. 
 The work of S.~K. is supported in part by Grant-in-Aid for Scientific Research on Innovative Areas, the Ministry of Education, Culture, Sports, Science and Technology, No. 16H06492 and No. 18H04587, Grant H2020-MSCA-RISE-2014 No. 645722 (Non-Minimal Higgs), and JSPS Joint Research Projects
(Collaboration, Open Partnership) “New Frontier of neutrino mass generation mechanisms via Higgs physics at LHC and flavor physics".

\end{acknowledgments}

\end{document}